\documentstyle[twoside,fleqn,epsfig,espcrc2]{article}
\newcommand{\msbar}{\overline{\mbox{\scriptsize MS}}}
\newcommand{\MSbar}{\overline{\mbox{MS}}}
\newcommand{\RI}{\mbox{RI}}
\newcommand{\ri}{\mbox{\scriptsize RI}}

\newcommand{\cals}{{\cal S}}

\newcommand{\Tr}{\mbox{Tr}\;}

\newcommand{\bc}{\begin{center}}
\newcommand{\ec}{\end{center}}
\newcommand{\be}{\begin{equation}}
\newcommand{\ee}{\end{equation}}
\newcommand{\bea}{\begin{eqnarray}}
\newcommand{\eea}{\end{eqnarray}}
\newcommand{\ba}{\begin{eqnarray}}
\newcommand{\ea}{\end{eqnarray}}
\newcommand{\brr}{\begin{array}}
\newcommand{\err}{\end{array}}

\def\dfrac#1#2{{\displaystyle {#1 \over #2}}}

\newcommand{\id}{\hbox{1$\!\!$1}}

\newcommand{\simge}{\ \lower-
1.2pt\vbox{\hbox{\rlap{$>$}\lower5pt
\vbox{\hbox{$\sim$}}}}\ }

\newcommand{\RGI}{\mbox{\scriptsize RGI}}
\newcommand{\Oa}{{\cal O}(a)}

\def\spose#1{\hbox to 0pt{#1\hss}}
\def\ltapprox{\mathrel{\spose{\lower 3pt\hbox{$\mathchar"218$}}
 \raise 2.0pt\hbox{$\mathchar"13C$}}}
\def\gtapprox{\mathrel{\spose{\lower 3pt\hbox{$\mathchar"218$}}
 \raise 2.0pt\hbox{$\mathchar"13E$}}}
\def\inapprox{\mathrel{\spose{\lower 3pt\hbox{$\mathchar"218$}}
 \raise 2.0pt\hbox{$\mathchar"232$}}}

\begin{document}

\title{Quenched Results for Light Quark Physics with Overlap Fermions
\thanks{Based on talks by L.~Giusti, C.~Hoelbling and C.~Rebbi.
Work supported in part under DOE grant DE-FG02-91ER40676.}}
\author{L.~Giusti\address{Centre de Physique Theorique CNRS Luminy, Case 907
F-13288 Marseille Cedex 9 France}, 
C.~Hoelbling\address{NIC/DESY Zeuthen, Platanenallee 6, D-15738 Zeuthen, Germany}, 
C.~Rebbi\address{Department of Physics - Boston University, 
590 Commonwealth Avenue, Boston MA 02215, USA}}

\begin{abstract}
We present results of a quenched QCD simulation with overlap fermions on
a lattice of volume $V = 16^3\times 32$ at $\beta=6.0$, which corresponds
to a lattice cutoff of $\simeq 2 $~GeV and an extension of $\simeq 1.4$~fm.
From the two-point correlation functions of bilinear operators we 
extract the pseudoscalar meson masses and the corresponding decay constants.
From the GMOR relation we determine the chiral condensate
and, by using the $K$-meson mass as experimental input, 
we compute the sum of the strange and average up-down quark masses  $(m_s + \hat m)$. 
The needed logarithmic divergent renormalization constant $Z_S$ is
computed with the RI/MOM non-perturbative renormalization technique.
Since the overlap preserves chiral symmetry at finite 
cutoff and volume, no divergent quark mass and chiral condensate 
additive renormalizations are required and the results are ${\cal O}(a)$ improved. 

\end{abstract}
\vspace{-0.5cm}
\maketitle
\section{Introduction}
In the last few years a major breakthrough in the lattice
regularization of Fermi fields was achieved through
the closely related domain wall \cite{DK} and overlap 
\cite{HN} formulations. Neuberger found 
a lattice Dirac operator $D$~\cite{neub1} which avoids
doubling and, most notably, satisfies
the Ginsparg-Wilson (GW) relation \cite{GW}.
Thus the corresponding action in the massless
limit preserves a lattice form of chiral symmetry  
at finite lattice spacing and volume 
\cite{neub1,luscher}. As a result the use of the 
overlap action entails many theoretical advantages 
(see ref.~\cite{reviews} and references therein for recent reviews);
in particular it forbids mixing among operators 
of different chirality and therefore it 
can be very helpful (if not crucial) for computing  
weak amplitudes.  No power-divergent subtractions
are needed to calculate the matrix elements relevant 
for the $\Delta I=1/2$ rule in $K\rightarrow \pi\pi$ 
decays~\cite{capgiu}. 

The theoretical advantages of overlap fermions come at the cost of an 
increased computational burden in numerical simulations. 
The main question we wish to answer in the work presented here is 
whether the overlap formalism can be effectively used for large scale 
QCD calculations, with known algorithms and the current generation of computers.
The calculation of light quark masses is an ideal test since it uses many 
of the ingredients needed for a generic phenomenological computation. 

We have performed a fully non-perturbative calculation of
$(m_s+\hat m)$ in the quenched approximation
following the procedure proposed in \cite{noi_mq}. 
We also report results for the chiral condensate
computed using  
the Gell-Mann--Oakes--Renner (GMOR) relation and for 
the pseudoscalar meson masses and decay constants. 
\section{Quark Masses and Chiral Condensate}\label{sec:Action}
The fermionic action in the overlap regularization reads 
\be\label{eq:overlap}
S = 
 \bar{\psi} \left[\Big(1-\frac{1}{2\rho}a M \Big)D + M\right] \, \psi
\label{eq:sg_QCD}
\ee
where $M$ is the matrix of bare masses $(m_1,m_2,\dots)$
in flavor space. The Neuberger-Dirac operator is defined as \cite{neub1}
\ba\label{eq:opneub}
D &=&  \frac{\rho}{a}\left( 1 + V \right) =
\frac{\rho}{a}\left( 1 + X\frac{1}{\sqrt{X^\dagger X}}\right)\\
X &=& D_W -\frac{1}{a}\rho\,
\ea
where $D_W$ is the Wilson-Dirac operator (definitions and 
conventions used here are fully described in 
ref.~\cite{ourselves}). 
The renormalized quark mass is defined as
\be
\hat{m}(\mu) = \lim_{a\rightarrow 0} Z_m(a \mu) m(a)
\ee
where $Z_m(a \mu)$ is a logarithmically divergent
renormalization constant and $m(a)$ is the bare mass 
parameter which appears in the action of eq.~(\ref{eq:overlap}) 
and in the corresponding axial and vector Ward identities.
The non-singlet ``local'' bilinear quark operators
we are interested in are defined as 
\be   
O_\Gamma(x) = \bar \psi_1(x) 
\Gamma \left[(1-\frac{a}{2\rho}D) \psi_2\right](x)   
\label{eq:o}   
\ee   
where $O_\Gamma  \equiv  \{V_{\mu}, A_{\mu},S,P\}$ correspond to
$\Gamma     \equiv \{\gamma_{\mu}, \gamma_{\mu}\gamma_5,\id,
\gamma_5\}$.
They are subject to multiplicative   
renormalization only, i.e.~the corresponding renormalized operators are
\be   
\hat O_\Gamma(x,\mu) = \lim_{a \rightarrow 0}   
\, Z_\Gamma(a \mu) O_\Gamma(x,a)
\label{eq:zodef}   
\ee   
where $Z_\Gamma(\mu)$ are the appropriate renormalization constants. 
Since $V_\mu$, $A_\mu$ 
and $S$, $P$ belong to the same chiral multiplets 
$Z_V=Z_A$ and $Z_S=Z_P$. Flavor symmetry imposes $Z_S = 1/Z_m$.

The bare chiral condensate is defined as
\be\label{eq:cond_def}
\chi(a) \equiv \lim_{m \rightarrow 0}\frac{1}{N_f}
\langle \bar \psi(0) [(1-\frac{a}{2\rho} D) \psi](0) \rangle
\ee
where $m$ in this case is a common mass given to the light quarks.
For non-zero quark mass 
\be\label{eq:cond_div}
\frac{1}{N_f}\langle \bar \psi(0) [(1-\frac{a}{2\rho} D) \psi](0) \rangle =
\chi(a) + \beta_\chi \frac{m(a)}{a^2}
\ee
since chiral symmetry forces the coefficient of the linear divergence to be zero.
The condensate satisfies the integrated non-singlet chiral Ward identity
\bea\label{eq:bella}
& & \frac{1}{N_f}
\langle \bar \psi(0) [(1-\frac{a}{2\rho} D) \psi](0) \rangle
= \\
& & - \, m \sum_{x} \langle P(x) \overline{P}(0)\rangle \nonumber
\eea
Therefore by writing the correlation function
$\langle P(x) \overline{P}(0) \rangle$ as a time-ordered product and by inserting a
complete set of states in standard fashion we can also write
\be
\label{eq:wip}
\chi(a) =
-\, \lim_{m \rightarrow 0}
\dfrac{m}{M_P^2}
\Big \vert \langle 0 \vert P \vert P \rangle \Big \vert^2
\ee
where $M_P$ is the mass of the pseudoscalar state $\vert P \rangle$.
If we use
\be\label{eq:fp}
2 m |\langle 0 | P | \pi\rangle| =  f_{P} M^2_{P}
\ee
where $f_{P}$ is the corresponding pseudoscalar decay constant, we arrive
to the familiar GMOR relation
\be\label{eq:GMOR_true}
\chi(a) =
- \lim_{m \rightarrow 0} \frac{f^2_{P} M^2_{P}}{4 m}
\ee
To preserve eq.~(\ref{eq:bella}), the renormalized chiral condensate is defined
as
\be
 \frac{1}{N_f}\langle \bar \psi \psi\rangle(\mu) = \lim_{a\rightarrow 0} Z_S(a \mu)
\chi(a)
\ee
\vspace{-0.5cm}
\section{Non-perturbative Renormalization}\label{sec:ZA} 
The bilinear renormalization constants $Z_\Gamma(a \mu)$
have been be computed at one loop in perturbation 
theory \cite{vicari,capgiu}. Nevertheless,
to avoid large uncertainties due to higher order terms, 
we prefer to compute them non-perturbatively.

The exact chiral symmetry implies
a conserved axial 
current~\cite{hasenfratz2,Kikukawa:1999py}
which enters the corresponding Ward identities. Therefore
the scheme and scale independent 
renormalization constant $Z_A$ can be computed from the 
axial Ward identity 
\be\label{eq:AWIloc}
Z_A \langle\bar\nabla_\mu A_\mu(x)  P(0) \rangle = 2\, m 
\langle P(x) P (0) \rangle + {\cal O}(a^2)
\ee
where $\bar\nabla_\mu$ is the symmetric lattice derivative.
To compute the logarithmic divergent 
renormalization constant $Z_S(\mu)$ we implemented 
the RI/MOM non-perturbative renormalization 
technique proposed in ref.~\cite{NP}. 
The amputated off-shell Green's functions are defined as 
\be   
\Lambda_\Gamma(p) = \cals ^{-1}(p) G_\Gamma(p) \cals ^{-1}(p)   
\label{eq:amp}   
\ee  
where $G_\Gamma(p)$ are the improved quark correlation functions 
and $\cals ^{-1}(p)$ is the improved (external) quark propagator in the 
Fourier space. The projected amputated Green's functions 
$\Gamma_\Gamma(p)$ are defined as   
\be
\Gamma_\Gamma(p) = \frac{1}{12} \Tr \left[P_\Gamma \Lambda_\Gamma(p)\right]   
\label{eq:proj_GF}   
\ee   
where the trace is over spin and color indices and $P_\Gamma$ are the Dirac matrices   
which renders the tree-level values of $\Gamma_\Gamma(p)$ equal to unity
(see \cite{ourselves} for more details).
$Z_S^{\ri}(a \mu)$  can be determined, up to ${\cal O} (a^2)$, 
by imposing the renormalization condition 
\be\label{eq:rimom}
Z^{\ri}_S(a \mu) =  \left. \lim_{m\rightarrow 0} Z_A \frac{\Gamma_A (p,m)}{\Gamma_S
    (p,m)}\right|_{p^2=\mu^2} 
\ee 
To compare the running of $Z^{\ri}_S(a \mu)$ with the evolution
predicted by the renormalization group equations, 
it is useful to define the renormalization 
group invariant (RGI) renormalization constant \cite{noi_np}  
\begin{equation}
\label{eq:Z_RGI} 
Z_S^{\RGI}(a)=\frac{Z^{\ri}_S(a \mu)}{c_S^{\ri}(\mu)}
\end{equation}
where $c_S^{\RI}(\mu)$ 
at the next-to-next-to-leading-order (NNLO)
is given in \cite{noi_np}.
Up to higher order terms in continuum perturbation theory
$Z_S^{\RGI}(a)$ is  independent of the renormalization scheme,
of the external states and is gauge invariant.
It is interesting to note the the non-degenerate scalar and pseudoscalar Green's 
functions satisfy the WI~\cite{G&V}
\ba\label{eq:moltobella}
& & (m_1 - m_2 ) \Gamma_S   \left(p,m_1,m_2\right)  = \\
& & m_1 \Gamma_P\left(p,m_1,m_1 \right) 
 -  m_2 \Gamma_P\left(p,m_2,m_2\right)   \nonumber
\ea
Equation~(\ref{eq:moltobella}) gives an exact relation among 
bare correlation functions of ``local''
scalar and pseudoscalar operators at non-zero quark masses, without
any reference to the conserved axial or vector currents.
\section{Numerical Results}
The numerical results we present
are based on a set of 54 quenched gauge 
configurations produced with the standard Wilson gauge action,
with $V = 16^3\times 32$ and $\beta=6.0$, which we
retrieved from the ``Gauge Connection'' repository~\cite{GaugeConn}. 
For each configuration 
we fixed the Landau gauge by requiring a quality factor of 
$\theta < 10^{-6}$ and calculated the quark propagators 
for $5$ different 
bare quark masses $m_i=\{0.040,0.055,0.070,0.085,0.0100\}$,
as described in appendix~\ref{numimp}. 
We computed then the
two-point correlation functions
\ba\label{eq:scal_vec}
G_{SS}(t) & = & \sum_{x}\langle S(x,t)\overline{S}(0,0) \rangle\\
G_{PP}(t) & = & \sum_{x}\langle P(x,t)\overline{P}(0,0) \rangle\\
G_{\nabla AP}(t) & = & \sum_{x}\langle \bar \nabla_0 A_0(x,t)
\overline{P}(0,0)\rangle
\ea
We estimated the errors by a jackknife procedure, blocking the data in
groups of three configurations, and we checked that blocking in groups
of different size did not produce relevant changes in the error estimates.
\subsection{Bare Quark Masses}
In the quenched approximation the contributions from chiral zero modes
is not suppressed by the fermionic determinant.
In particular $G_{PP}(t)$ receives 
unsuppressed contributions
proportional to $1/m^2$ and $1/m$,
which should vanish in the
infinite volume limit
but can be quite sizeable at finite volume \cite{BNL_spec}.
These quenching artifacts cancel in the difference 
of pseudoscalar and scalar meson propagators \cite{BNL_spec}
\be\label{eq:BNL_trick}
G_{P-S}(t) = G_{PP}(t) - G_{SS}(t)
\ee
The drawback is, of course, that the plateau
in the effective mass and, correspondingly, the range that can
be used for the $\cosh$ fit become shorter 
(See~\cite{ourselves} for more details).
\begin{table}[htb]
\begin{center}
\begin{tabular}{cccc}
\hline
$a m $ & \multicolumn{3}{c}{$G_{P-S}$} \\
\hline
        & $Z_{P-S}$ & $a M_{P}$ & $a f_{P}$ \\
\hline
0.100   & 0.0040(4) & 0.379(6)  & 0.089(2)  \\
0.085   & 0.0036(4) & 0.348(6)  & 0.085(2)  \\
0.070   & 0.0033(4) & 0.315(7)  & 0.081(2)  \\
0.055   & 0.0030(5) & 0.280(9)  & 0.076(2)  \\
0.040   & 0.0026(5) & 0.239(11) & 0.071(2)  \\
\hline
\end{tabular}
\caption{Meson masses and pseudoscalar matrix elements for all the bare quark masses
considered in the simulations, as obtained from $G_{P-S}(t)$.
\label{tab:masses}}
\end{center}
\end{table}
We fitted 
$G_{PP}(t)$ and $G_{P-S}(t)$ to a single particle propagator 
in the time intervals $t_1-t_2=12-16$ and $10-16$ respectively. 
The two correlation functions give values for the pseudoscalar masses compatible 
within the statistical errors. One does not notice 
any sign of the singular contributions from zero modes
in the masses obtained
from the pseudoscalar correlation function.  These are expected to show up
at some point, but one would probably need much higher
statistical accuracy and lower values of $m$ to bring them into evidence. 
The results for the matrix elements, parameterised by the factors $Z_{S-P}$ and 
$Z_{PP}$ for $G_{S-P}(t)$ and $G_{PP}(t)$, respectively, 
show more significant differences.
Within the large statistical errors, $Z_{S-P}$ and 
$Z_{PP}$ are still compatible. However the central 
values are quite different and the fact that a curvature 
(see \cite{ourselves} for more details) shows up only
in the results for $Z_{PP}$
points to the fact that what we are seeing is the effect of 
the unsuppressed zero modes,
and not of chiral logarithms which would affect both sets
of results (and would most likely become noticeable at much smaller
values of $a m$).  On account of the above, we report in 
table~\ref{tab:masses} the results obtained from $G_{S-P}(t)$ 
which we will use to derive our further results.  It must also be said
that most of the observables will be calculated directly at
$m\simeq m_s/2$ (see below), and for these
the difference between $Z_{PP}$ and $Z_{S-P}$ is irrelevant within
statistical errors.
\begin{figure}[htb]
\begin{center}
\epsfig{file=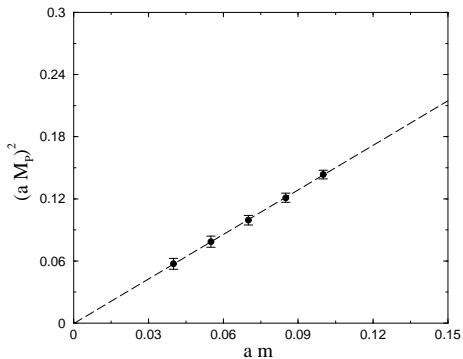,width=6.0cm}
\vspace{-1.0cm}
\caption{$(a M_{P})^2$ vs.~$a m$ as obtained from $G_{P-S}(t)$. 
The dashed lines represent the results of a linear fit.
\label{fig:mass_extrapo1}}
\end{center}
\end{figure}
We illustrate in fig.~\ref{fig:mass_extrapo1} the
values for $(a M_{P})^2$, obtained from $G_{P-S}(t)$, as a
function of the bare quark mass $a m$. 
A linear behaviour
\be
(a M_{P})^2 = {\cal A}_{M_P} + {\cal B}_{M_P} (a m)
\ee
fits very well the data with
\be
{\cal A}_{M_P} = -0.0005(68)  \qquad  {\cal B}_{M_P} = 1.43(7)
\label{eq:mps_fit2}
\ee
The vanishing of the intercept within statistical errors 
signals the absence of additive quark mass renormalization.
\begin{figure*}[htb]
\vspace{-1.0cm}
\newlength{\digitwidth} \settowidth{\digitwidth}{\rm 0}
\catcode`?=\active \def?{\kern\digitwidth}
\begin{center}
\begin{tabular}{cc}
\mbox{\epsfig{file=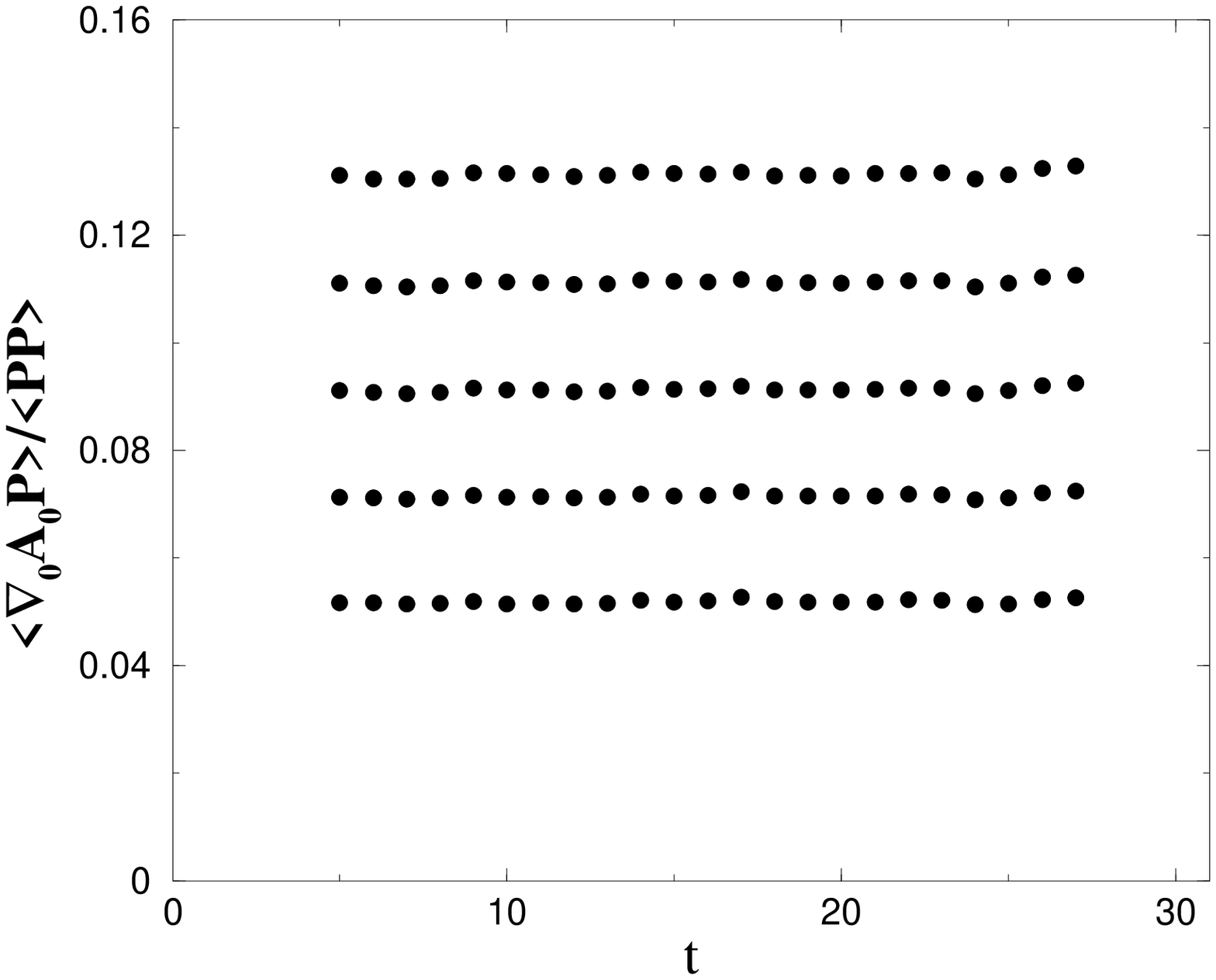,width=6.0cm}} &
\mbox{\epsfig{file=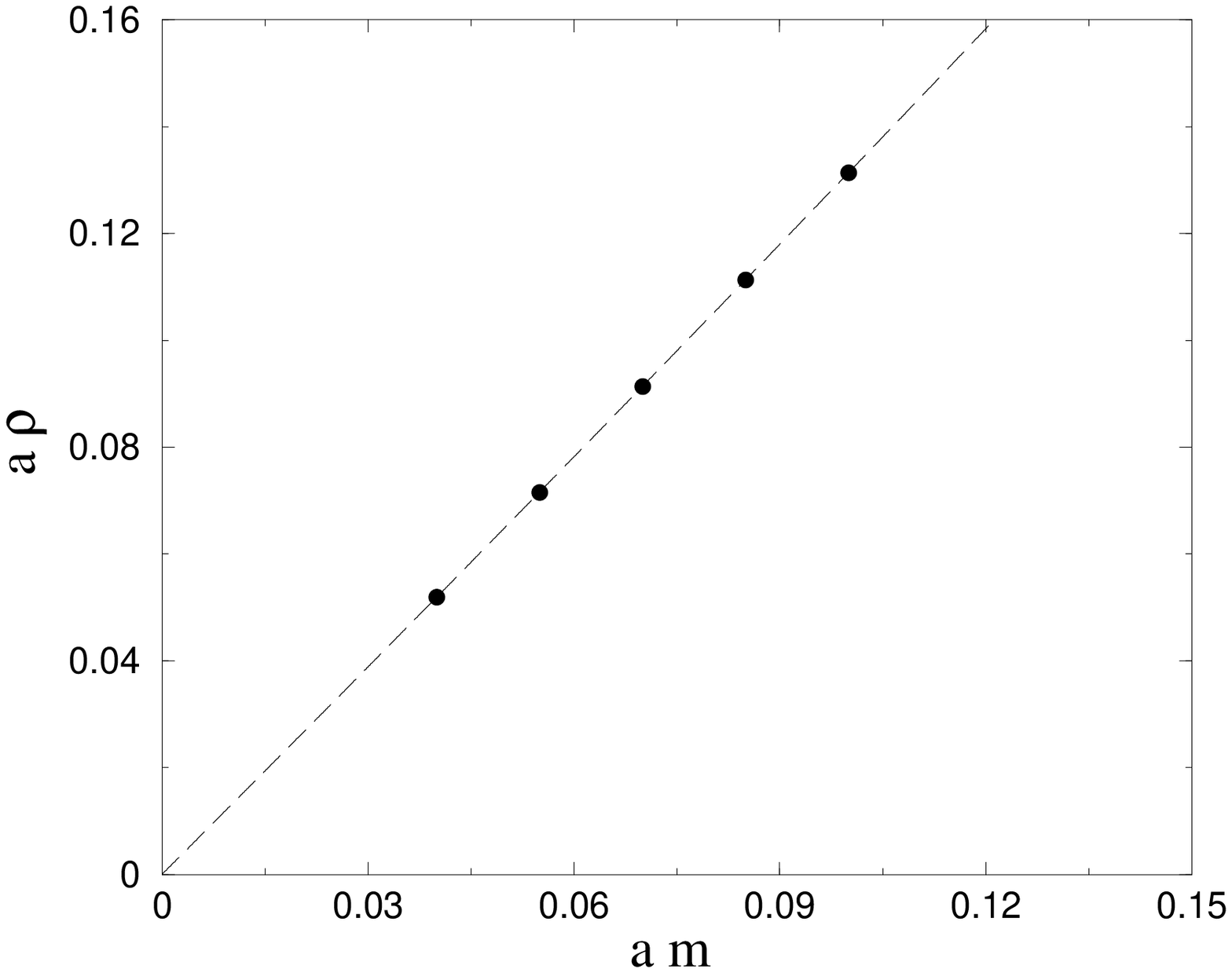,width=6.0cm}} \\
\end{tabular}
\vspace{-1.0cm}
\caption{Left: $G_{\nabla AP}/G_{PP}$ vs.~$t$ for all simulated masses. 
Right: $(a \rho)$ as a function of the 
bare quark mass.  The dashed line
represents the result of a quadratic fit.
\label{fig:axials}}
\end{center}
\end{figure*}
In order to fix the lattice spacing, we used the method of
``lattice physical planes'' \cite{lp-method}, in which the ratio
$f_K/M_K$ is fixed to its experimental value.
This avoids recourse to a chiral extrapolation for observables
except when it is really needed.
For more details we refer the reader to \cite{ourselves}
and only quote here the result 
\be
a M_K = 0.216(8)  \qquad a f_K = 0.0698(26)
\label{eq:amk}
\ee
To fix the lattice spacing we compared the
value of $a f_K$ with its experimental value and we obtained
$a^{-1}_{f_K} = 2.29(9)$
We also computed the chiral condensate by
using the GMOR relation. The quantity
\be
a^3 \chi_m = -\, (a m) \frac{Z_{P-S}}{(a M_P)^2} 
\ee
exhibits a very good linear behaviour in the bare quark mass and
a linear fit leads to 
\be\label{eq:gmor_numerics}
a^3 \chi = -\, 0.00117(27)
\ee
The chiral condensate can also be computed directly from its definition
in eq.~(\ref{eq:cond_def}). Using again a computational strategy
similar to that used in eq.~(\ref{eq:BNL_trick}) in order
to take care of the infrared divergent 
contributions (in $m$) from unsuppressed zero modes, we obtain
\be
a^3 \chi(a) = -\, 0.00117(42)
\label{eq:chidirect}
\ee
which is in remarkable agreement with eq.~(\ref{eq:gmor_numerics}).
\subsection{Renormalization of the Axial Current}
We implemented eq.~(\ref{eq:AWIloc}) numerically by 
computing the ratio
\be
\label{eq:rat_AWI}
R_\rho(t) = \frac{G_{\nabla AP}(t)}{G_{PP}(t)}
\ee
The results obtained for each mass are reported in
the first plot of fig.~\ref{fig:axials}.
The flatness of the data highlights the improvement 
achieved with the overlap regularization. We fitted $R_\rho(t)$
to a constant $\rho(m)$ in the range $5 < t < 27$ and then performed 
a quadratic fit (second plot in fig.~\ref{fig:axials})
\be
\rho(m) = A_\rho + B_\rho m + C_\rho m^2
\ee
obtaining
\bea
A_\rho &=& -0.00002(7) \qquad B_\rho = 1.286(3) \\
C_\rho &=& 0.277(12)  \nonumber
\eea 
Note that the intercept is compatible with zero:
this should not come as a surprise
since $A_\mu(x)$ has the correct behaviour under 
global non-singlet chiral transformations.

The axial current renormalization constant $Z_A$
is given by $Z_A=\frac{2}{B_\rho}$.
We also fitted $\rho(m)$ linearly in 
the quark mass and we take the difference 
of the quadratic and linear results as a rough 
estimate of the systematic error. Our final result 
is $Z_A=1.55(4)$ \cite{ourselves}.
This value is larger than the one obtained in refs.~\cite{vicari,capgiu} 
using standard lattice perturbation theory at one-loop.
\begin{figure*}[htb]
\vspace{-1.0cm}
\catcode`?=\active \def?{\kern\digitwidth}
\begin{center}
\begin{tabular}{cc}
\mbox{\epsfig{file=Gs_Gp_wiq.eps,width=6.0cm,height=5.0cm}} &
\mbox{\epsfig{file=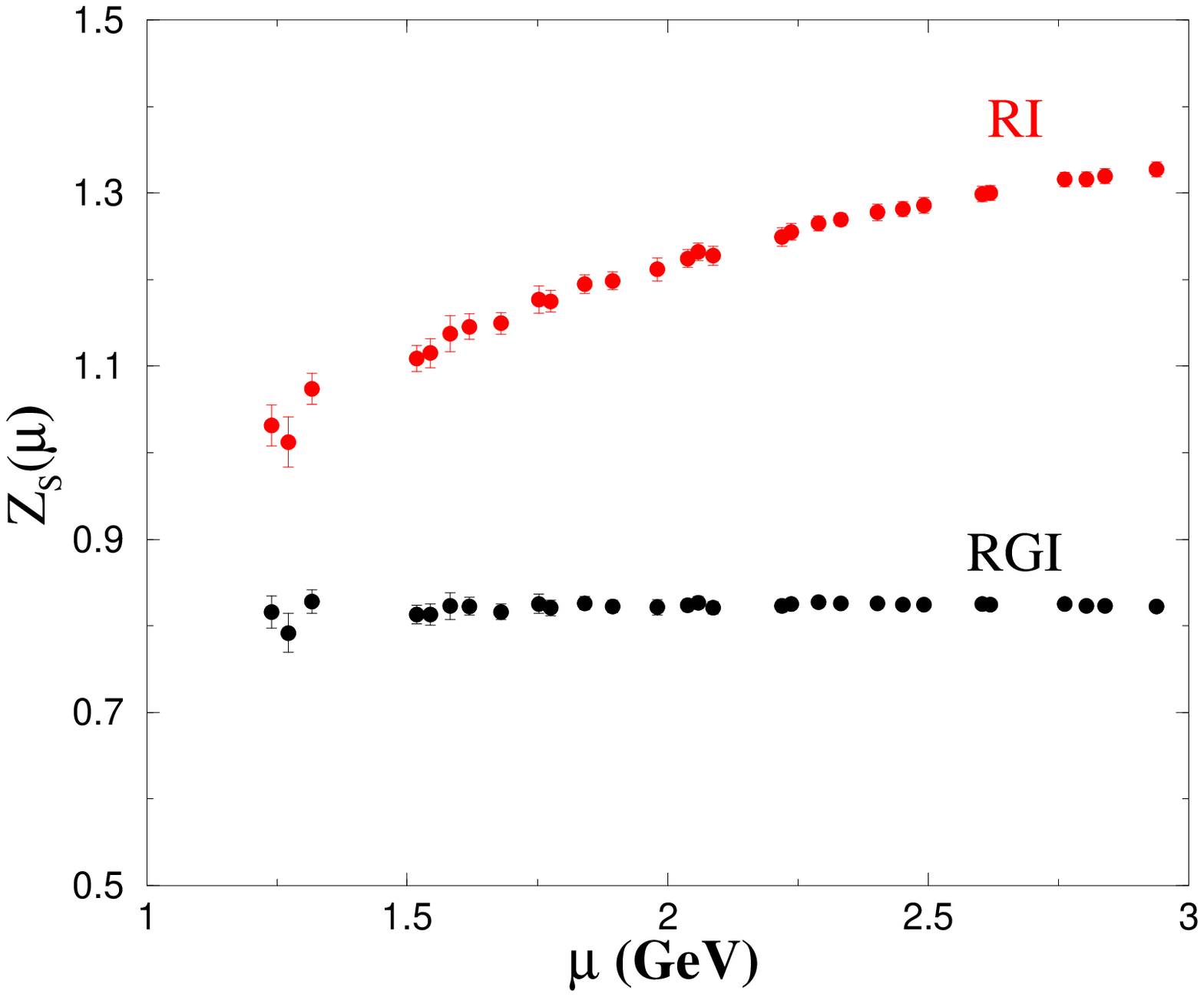,width=6.0cm}} \\
\end{tabular}
\vspace{-1.0cm}
\caption{$R_{PS}$ (left) and $Z_S(a\mu)$ (right) as a function of 
the renormalization scale.
\label{fig:moltobella}}
\end{center}
\end{figure*}
\subsection{Renormalization of the Scalar Density}
To implement the RI/MOM non-perturbative renormalization technique
we computed the projected, amputated Green's functions
of the quark bilinears defined in eq.~(\ref{eq:proj_GF}). 
In the first plot of fig.~\ref{fig:moltobella} we show
the ratio
\be
R_{PS} = \frac{(m_1 - m_2 ) \Gamma_S   
\left(p,m_1,m_2\right)}{m_1 \Gamma_P\left(p,m_1,m_1 \right) -
m_2 \Gamma_P\left(p,m_2,m_2\right)} 
\ee
for a given combination of masses. As expected from (\ref{eq:moltobella})
its value is always compatible with one. The ratio of the vector and 
axial Green's functions produces analogous results which we do not 
report here for lack of space. On the other hand the 
ratio $\Gamma_P/\Gamma_S$ with degenerate masses
is expected to be sensitive to non-perturbative 
Goldstone pole contributions \cite{politzer,NP}. 
A detailed analysis will be presented in a forthcoming paper. 

To calculate $Z_S$ in the RI/MOM scheme, 
we computed the ratio \cite{ourselves}
\be
R_{AS}(m)=Z_A\frac{\Gamma_A(p,m)}{\Gamma_S(p,m)}
\ee
which we extrapolated to the chiral limit as
\be
R_{AS}(m)=Z_S^{RI}+A m^2+B m^{-2}
\ee
to take into account the effect of unsuppressed zero modes.
The results we thus obtained, shown in the second plot of 
fig.~\ref{fig:moltobella}, lead to 
\be
Z_S^{\ri}(2\, \mbox{GeV}) = 1.24(5) 
\label{eq:zs}
\ee
where the error is mainly systematics and is 
due to the uncertainty in the value of the renormalization scale
and to the chiral extrapolation (the latter estimated
by using different forms of the extrapolating function).
Using known results from NNLO continuum perturbation theory\cite{noi_np},
we convert eq.~(\ref{eq:zs})
to the $\MSbar$ scheme, obtaining
\be\label{eq:zs_MSBAR}
Z_S^{\msbar}(2\mbox{GeV})=1.41(6)
\ee
This result is in good agreement with the one 
in~\cite{HJLW} which was determined by fixing 
the renormalized quark mass value.
Invoking again NNLO continuum perturbation theory, 
we calculate $Z_S^{RGI}$ from eq.~(\ref{eq:Z_RGI}). We use $N_f=0$ and 
$\Lambda_{QCD}=0.238(19)$~\cite{lambdaqcd}. The result, 
displayed in the second plot of fig.~\ref{fig:moltobella}
shows the remarkable flatness of 
$Z_S^{RGI}$ for $\mu>1.5$~GeV which 
confirms the high level of improvement reached by the overlap regularization.
Averaging over the region $(1.5-3)$~GeV, we find
\be
Z_S^{RGI}=0.82(3)
\ee
\vspace{-0.5cm}
\section{Physical Results}\label{sec:Results}
In this section we will use our lattice results to infer the
renormalized values of the sum of the strange and
average up-down quark masses and of the chiral condensate.
From eq.~(\ref{eq:amk}) we obtain
\be
m_s(a) + \hat m(a) = 149(9)\, \mbox{MeV}
\label{eq:msbare}
\ee 
where the error is statistical only. Since our volume is fairly large,
we expect our main sources of systematic errors to come from 
discretization effects and from the quenched 
approximation. For a rough estimate of the systematic error due to 
quenching approximation we can use the results in~\cite{CPPACS_nucleon},  
from which one sees that, within the quenched approximation, 
the use of alternative observables to 
calculate the lattice spacing can produce differences 
of approximately $10\%$.  Had we used $r_0$ \cite{Guagnelli:1998ud} 
to fix our lattice spacing   
we would have obtained a number $\sim 7\%$ higher than the one in
eq.~(\ref{eq:msbare}).  Combining this fact with the
results in~\cite{Garden:2000fg}, we infer 
that discretization uncertainties are below $10 \%$.  
If we had used the experimental NNLO results
from $\alpha_s(M_Z) = 0.118$ in the matching of the renormalization constants
in eq.~(\ref{eq:zs_MSBAR}), we would have obtained
a value of the scalar renormalization constant $\sim 10\%$
higher than the one given above.  The difference can be taken as an 
indication of the systematic error introduced by the quenched approximation.
In order to be conservative, we will take $15 \%$ as the estimate
of our overall systematic error in the renormalized quark masses 
due to quenching and discretization effects. A more precise estimate 
of the systematic errors will need much more extensive simulations, 
which would go beyond the capability of our current computer resources
and the exploratory nature of the our work.
Combining the results in eqs.~(\ref{eq:zs}) and~(\ref{eq:msbare}) we obtain
\be
(m_s + \hat m)^{\ri}(\mbox{2 GeV}) = 120 \pm 7 \pm 21\, \mbox{MeV} 
\label{eq:ml+ms}
\ee
which represent one of the main results of this work.
By using the quark mass ratio $m_s/\hat m =24.4 \pm 1.5$ from chiral perturbation theory
\cite{Leutwyler} the above translates to 
\be
m_s^{\msbar}(\mbox{2 GeV}) = 102 \pm 6 \pm 18 \, \mbox{MeV} 
\ee
This result agrees very well with the current lattice world 
average \cite{vittorio_rev}. 

Insofar as the value of the condensate is concerned,
if we used the standard two-step approach, i.e.~first measure the 
dimensionless condensate, see eqs.~(\ref{eq:gmor_numerics}) 
or~(\ref{eq:chidirect}), and then
multiply it by the cubic power of the lattice spacing,  
the result would be affected by a very large 
systematic error due to the uncertainty in the determination of the 
lattice spacing in quenched simulations. 
Instead, we will use an alternative method \cite{noi_cond}.
We write the GMOR relation (\ref{eq:GMOR_true})
for the renormalized condensate as follows
\be
\chi(a)
= - \dfrac{1}{4} f_\chi^2 {\cal B}_{M_P} a^{-1}
\label{eq:psi1}
\ee
where ${\cal B}_{M_P}$ is defined in eq.~(\ref{eq:mps_fit2}) and
$f_\chi=0.1282$~GeV is the ``experimental'' value, in physical units,
of the pseudoscalar decay constant extrapolated to the chiral limit.
While computing the condensate from the
above formula relies on an additional element of experimental
information, it has several advantages. The most important is that, by
expressing the condensate in terms of $f_\chi$, 
we are left with only one power of
the UV cutoff $a^{-1}$.
With this method we obtain
\bea
\label{eq:cond_best}
&&\langle\bar \psi \psi\rangle^{\ri}(\mbox{2 GeV}) = \\
&& -\, 0.0167 \pm 0.0010 \pm
0.0029 \, \mbox{GeV}^3 \nonumber
\eea
where the estimate of the systematic error has been made using the
same criteria we used to compute the error for the quark masses.
This is our best value for the chiral condensate.
It is interesting to note that, if we had used the standard technique,
starting from eq.~(\ref{eq:gmor_numerics}), we would have obtained 
a result with a central value very close to the value in eq.~(\ref{eq:cond_best}),
but with a much higher error.
Finally, from eq.~(\ref{eq:cond_best}) and NNLO matching, 
we get
\bea
\label{eq:cond_msbar}
\langle\bar \psi \psi\rangle^{\msbar}(\mbox{2 GeV}) = -\, \left(267 
\pm 5 \pm 15 \, \mbox{MeV}\right)^3 
\eea
This result is in very good agreement with the result obtained by the authors 
of refs.~\cite{HJL_NP2,hasnf}, while it is smaller than the result 
in \cite{GrandDe}, even if still compatible within 
errors. Our result is also compatible
within errors with the number obtained few years ago 
in \cite{noi_cond} with Wilson-type fermions. 
We expect, though,
the systematics due to the discretization effects to be smaller 
(${\cal O}(a^2)$) in the result reported in 
eq.~(\ref{eq:cond_msbar}) than 
the error ($\Oa$) which affects the determination 
in \cite{noi_cond}.  
\section{Conclusions}\label{sec:conclusions}
At this conference we presented results for the 
pseudoscalar meson masses and decay constants 
together with the first fully non-perturbative computation
of $(m_s+\hat m)$ with overlap fermions in the quenched approximation.
We also computed the chiral condensate $\langle \bar \psi \psi \rangle$
from the GMOR relation and directly.
To avoid uncertainties due to lattice perturbation
theory, we computed the multiplicative renormalization constant 
$Z_S(\mu a)$ non-perturbatively in the RI/MOM scheme.
While the systematics errors due to quenching are common
to previous calculations, the other systematic errors
(mostly discretization effects) are different than in other lattice
regularizations and likely to be smaller, because of chiral symmetry.
Our results have indeed 
produced a remarkable verification of ``good chiral behavior''
of the overlap fermions
both in the axial Ward identity and for the pseudoscalar masses.

The calculation of light quark masses uses many of the ingredients 
needed for a lattice calculation of weak matrix elements, although
the latter is computationally more demanding.  From this point
of view, the very good agreement between our results for the quark
masses and the current lattice world average
bodes well for the use of the overlap formalism also in matrix
element calculations.  Our investigation has been
mostly of exploratory nature.  One would need to extend it to
larger volumes and better statistics.  
Nevertheless, we believe that it demonstrated that 
the overlap formalism can be used effectively, with known algorithms and the
present generation of computers, for large scale QCD calculations, at
least in the quenched approximation.  Thus we would conclude that
it represents a very promising non-perturbative
regularization for solving long standing problems, such as the proof
of the $\Delta I =1/2$ rule and the calculation of
$\epsilon'/\epsilon$.
\appendix
\section{Implementation of the Overlap Operator}\label{numimp}
The complicated form of Neuberger operator 
renders its numerical implementation more involved and expensive than those of the 
most common regularizations. The crucial point is to implement
the sign function
\be
\epsilon(H)  =  \mbox{sign}(H)
\ee
over the whole spectrum of eigenvalues of $H = \gamma_5 X$.
Exact diagonalization is feasible \cite{GHR_noi} for two dimensional 
models, but becomes prohibitive for larger systems such as QCD. 
Many algorithms have been proposed
in the literature \cite{Neuberger:1998my,Hernandez:1999et,Edwards:1999yw}. 
We opted for the optimal rational function approximation 
suggested in refs.~\cite{Neuberger:1998my,Edwards:1999yw}.
Starting from the observation \cite{Neuberger:1998my} that the sign function 
can be written as 
\ba
\varepsilon(x) & = & \lim_{N\rightarrow\infty} \varepsilon_N(x) \nonumber\\
\varepsilon_N(x)  & = & x \left(c_0 + \sum_{k=1}^{N} \frac{c_k}{x^2 +
    q_k}\right)
\label{eq:beauty_N}
\ea
a good approximation can be obtained at finite $N$ by fixing the coefficients
$c_k$ and $q_k$ with the Remez algorithm \cite{Edwards:1999yw,Remez}. 

The numerical procedure we have followed to obtain a 
quark propagator from a fixed source vector $|\eta\rangle$, i.e. solving the equation
\be\label{eq:difficult}
\left[(1-\frac{1}{2\rho} m_i )D + m_i\right]|\chi_i\rangle = |\eta\rangle
\ee
is the following:
\begin{itemize}
\item We have extracted the $15$ smallest eigenvalues of $H^2$ 
and the corresponding eigenvectors 
minimizing  the Ritz functional~\cite{Kalkreuter:1996mm}, projected
them out and treated the
action of $\epsilon(H)$ on this subspace exactly. 
The minimum precision required for each eigenvalue $\lambda^2_i$ is 
\be
r_{ritz} \equiv  \sqrt{\frac{|H^2|\lambda_i\rangle - \lambda^2_i|\lambda_i\rangle|^2  }
{\langle \lambda_i |\lambda_i\rangle }} < 10^{-6}
\ee  
where $|\lambda_i\rangle$ is the corresponding eigenvector. We checked on all the 
configurations that the largest eigenvalue extracted is always $|\lambda_{max}|>0.15$.
\item We have rescaled the reduced matrix $\tilde H = H/r$ by a factor
$r=3.7$ and we have approximated $\epsilon(\tilde H)$ as in eq.~(\ref{eq:beauty_N})
with  $N=14$ and $c_k,q_k$ fixed with the Remez algorithm 
\cite{Edwards:1999yw,Remez} which guarantees a precision 
\begin{equation}\label{eq:best_bound}
|\epsilon(\tilde H)-\epsilon_{N}(\tilde H)|\le 3.6\times 10^{-6}
\end{equation}
in the range $\tilde H \in [0.040,1.8]$.
In order to invert all the equations 
\begin{equation}\label{eq:in_inversion}
\left(\tilde H^2 + q_k \right)|\psi_k\rangle=|s\rangle \qquad  N=1,\dots 14
\end{equation}
in one stroke we use an (inner) multi conjugate gradient (MCG) 
solver \cite{Jegerlehner:1996pm}. 
Note, that during the inner MCG it is 
necessary to store only $N+2$ large vectors 
since we are only interested in
\begin{equation}
|\psi\rangle=\sum_{k=1}^{N}c_k|\psi_k\rangle\
\end{equation}
The convergence is governed by the inversion corresponding to the 
smallest $q_k$ for which we require a residual 
\be\label{eq:sofferenza}
r_{in} \equiv  \sqrt{\frac{|(\tilde H^2 + q_{14}) |\psi_{14}\rangle - |s\rangle|^2}
{\langle s |s\rangle}} < 10^{-6}
\ee 
Notice that having extracted the lowest eigenvectors of $H^2$ improves 
the approximation used for $\varepsilon(H)$ and reduces the condition number 
of $H^2$, which results in a speed up of the inversion in eq.~(\ref{eq:in_inversion}).
\item From eq.~(\ref{eq:difficult}) and using the GWR 
we can write
\be
|\chi_i\rangle = \frac{\rho\left((1-\frac{m_i}{2\rho})D^{\dagger} + 
m_i\right)}{\left(\rho^2 - m_i^2/4\right)\left(D + D^{\dagger}\right) + \rho m_i^2} |\eta\rangle
\ee
and therefore we are left to invert
\be
\left(D + D^{\dagger} + \frac{\rho m_i^2}{\rho^2-m_i^2/4}\right) |\tilde
\chi_i\rangle
= |\eta\rangle
\ee
$D+D^\dagger$ is an Hermitian matrix and therefore a MCG can be applied also 
in this case. Moreover $[D+D^{+},\gamma_5] = 0$ implies that we can always use
sources $|\eta\rangle$ and solutions $|\chi_i\rangle$ restricted 
to one chiral sector, saving one application of $\varepsilon(H)$ per 
iteration \cite{Edwards:1999yw}. 
The convergence is governed by the inversion corresponding to the 
smallest quark mass, for which we require a residual of
$r_{out} < 10^{-5}$ 
\end{itemize}
We did not attempt to use any preconditioning for the inner and the outer
MCG. At the $i$-th step of the outer MCG, the difference 
$|d_{i}\rangle$
of two consecutive source vectors $|s_{i}\rangle$ and $|s_{i-1}\rangle$
on which $\varepsilon_N(H)$ has to be 
applied gets smaller and smaller when $r_{out}$ decreases. 
Therefore at the $i+1$ step we have applied $\varepsilon_N(H)$ directly 
on $|d_{i}\rangle$ with an increased $r_{in}$ 
(every $40$ outer MCG step we required always the full precision to 
avoid rounding accumulation).  
After convergence, we always checked that the true residual 
associated with the eq.~(\ref{eq:difficult}) for each mass is $<10^{-5}$.
In our simulations the average numbers of inner and outer iterations is  
${\cal O}(150)$ and ${\cal O}(250)$ respectively.

\end{document}